\documentclass[preprint,11pt,authoryear, xcolor=table]{elsarticle}

\usepackage[misc]{ifsym}

\usepackage[a4paper, total={6in, 10in}]{geometry} 
\usepackage{apalike}
\usepackage{natbib}
\usepackage{booktabs,array}
\usepackage{graphicx}
\usepackage{multicol}
\usepackage{setspace}
\usepackage{appendix}
\usepackage{amsmath}
\usepackage{url}
\usepackage{pdfpages}
\usepackage[skip=0.5ex]{subcaption}
\usepackage[singlelinecheck=false]{caption}
\usepackage{listings}
\usepackage{xcolor,svgcolor}
\usepackage[plainpages=true,verbose]{hyperref}
\makeatletter
\hypersetup{
breaklinks=true,
colorlinks=true,
 linkcolor=darkgreen,
 citecolor=darkgreen,
 urlcolor=darkred,
}

\makeatother

\author[affil1,affil2]{ \Letter  Florentina Șoiman}
\author[affil2,affil3]{Mathis Mourey}
\author[affil1]{Jean-Guillaume Dumas}
\author[affil2]{Sonia Jimenez-Garces}
\address[affil1]{Univ.\ Grenoble Alpes, UMR CNRS 5224, LJK, 38000 Grenoble, France}
\address[affil2]{Univ. Grenoble Alpes, Grenoble INP, CERAG, 38000 Grenoble France}
\address[affil3]{The Hague University of Applied Sciences, Netherlands }
\address{\textit{[Firstname.Lastname]@univ-grenoble-alpes.fr}}

\begin{document}
\begin{frontmatter}
\title{\boldmath The forking effect}

\begin{abstract}
This study introduces the concept of {\em the forking effect} in the cryptocurrency market, specifically focusing on the impact of forking events on bitcoin, also called parent coin. We use a modified exponential GARCH model to examine the bitcoin’s response in returns and volatility. Our findings reveal that forking events do not significantly affect the bitcoin’s returns but have a strong positive impact on its volatility, especially when considering market dynamics. Our model accounts for key features like volatility clustering and fat-tailed distributions. Additionally, we observe that following a fork event, volatility remains elevated for the next three days, regardless of other forking events, and the volatility impact does not increase when multiple forks occur simultaneously on the same day.\\
\end{abstract}


\pagebreak
\end{frontmatter}

\newpage
\section{Introduction}
The rapid advancement of Blockchain technology poses ongoing challenges for researchers and professionals globally. Its intricate nature and wide-ranging implications often lead to misconceptions, compounded by a lack of understanding about Blockchain, leading to irrational behavior and, consequently, market inefficiencies \citep{Dumas2021, Aste2019}. These arguments could explain why professionals from various fields (engineers, economists, regulators, etc.) are keen to enlighten the 'complicated' crypto world and propel its development.\\

With this study, we propose a research on the causal link between pure technological events, namely forks, and the cryptocurrency's financial characteristics. We bring to light the {\em forking effect}, which is the financial impact experienced by a cryptocurrency when forking events happen\footnote{A Blockchain fork is a modification, a discrepancy, or a breach in the consensus protocol, which can lead to a chain split and sometimes (when it's a hard fork) to the creation of a new cryptocurrency.}. Despite recent efforts to enrich the literature on cryptocurrencies, we observed a general lack of financial research on the topic of Blockchain forks. This study aims to bridge this gap by addressing the following research question: \textit{How do bitcoin's financial characteristics react to forking events?}. Bitcoin is the most well-known and forked cryptocurrency. This research exclusively examines the bitcoin forking events; therefore, even though our `parent coin' will always be bitcoin, we will continue to refer to it in a general manner, establishing in this way a theoretical concept that could be further applied when analyzing the forks of other cyptocurrencies. Our sample accounts for 93 bitcoin forks that occurred between 2014 and 2020. In our observations, we have noticed that forking events in the crypto-market frequently happen on the same day or in close succession, rendering the traditional event study methodology \citep{MacKinlay1997} impractical for studying these events. Therefore, to address our research question effectively, we will adopt a similar methodology to the one employed by \cite{Grobys2021} and use a modified version of the exponential GARCH (EGARCH) model. \\

In this analysis, we conduct our model estimation twice: firstly, following a similar approach to \cite{Grobys2021} without considering the market factor, and secondly, incorporating the market dynamics by considering the CRIX index. The motivation behind this choice is as follows. As we will show in the subsequent sections, the majority of forking events occur during the Bitcoin Bubble period. Given the significance of market risk premium as a predictor of cryptocurrency returns \citep{dunbar}, we deemed it relevant to explore any potential differences between these two models. Our findings indicate that forking events do not affect the returns of the parent coin on the day of occurrence, a result consistent across both models. However, we observe a strong impact of forks on bitcoin's volatility, particularly when considering market dynamics. In addition, our investigations into multiple forks on the same day and consecutive-day forks demonstrate that the uncertainty generated by a fork does not intensify with simultaneous forks, and volatility remains elevated for three subsequent days after a fork event, regardless of other events. Furthermore, our robustness checks confirm the reliability and validity of our findings, as consistent results were obtained when analyzing all forking events together or exclusively focusing on hard forks.\\

Our study stands apart from previous research by being the first to investigate the impact of forking events and introducing the novel approach of employing an EGARCH model to assess their effects. This paper contributes to the understanding of Blockchain forks from both technological and financial points of view. Our findings hold significant implications for crypto-investors, emphasizing the importance of considering the impact of technological events in order to effectively manage risks within this market.\\

The following section exposes the research background comprising the description of Blockchain forks' characteristics and the development of the research question. Section 3 outlines the data and methodology employed, including the measures utilized. Section 4 delves into the results obtained and examines their implications. Section 5 outlines the supplementary tests conducted. Lastly, Section 6 concludes the study, presents potential avenues for future research, and acknowledges the limitations of this work.

\section{Research background}
In this section, we offer an overview of Blockchain forks and review the relevant literature on this topic. Then, we outline the paper's contribution and present the research question.

\subsection{Understanding Blockchain forks}
Cryptocurrencies are programmed/digital coins that do not exist in physical form and use Blockchain technology for operational purposes. Blockchain technology, a variant of distributed ledger technology (DLT), functions as a decentralized database. It operates by organizing transactional data into blocks, which are subsequently interconnected to form a chain-like structure.
Compared to national currencies, cryptocurrencies' operations are performed in a decentralized way. In this context, there is no longer a central point of control, such as traditional banks. Instead, every participant within a cryptocurrency network possesses access to the complete transactional data history and can actively contribute to the validation process \citep{Olleros2016, Button2019}.
Among many aspects that differentiate cryptocurrencies, an important one is the consensus protocol used by the Blockchain technology. This algorithm works as a manager for the entire database. More specifically, the consensus protocol is \textit{responsible} for the Blockchain's decentralization function; it enables the participants to engage in the validation process, assuring the majority's agreement on a unified transaction ledger \citep{Xiao2020}.

\subsubsection{What is a fork?}
\begin{figure}[h!]
\caption{\label{Fig1}\textbf{Forks' classification} \\ \footnotesize{\textit{Schematic representation of forks classification.}}}
\centering
\includegraphics[scale=0.8]{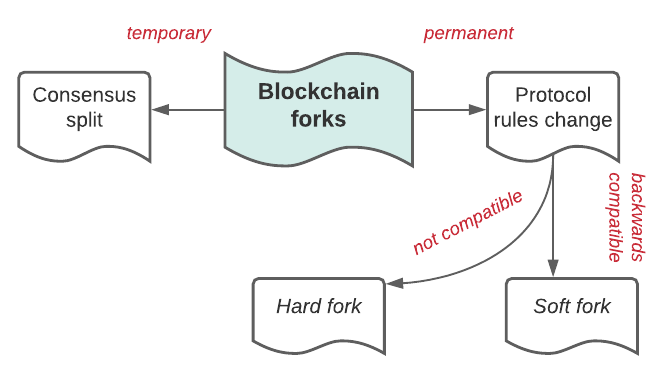}
\end{figure}

In the Blockchain world, a fork represents a modification, a discrepancy, or a breach of its consensus protocol. Similar to, for example, our computers' OS software that makes updates and upgrades all the time, the Blockchain consensus algorithm needs to evolve and undergo regular changes \citep{Islam2019b}. Often, Blockchain forks are acknowledged as exclusive chain splits, however, this is not always the case. Sometimes, the consensus protocol is modified while the chain structure remains intact \citep{BitMEX2017}. In figure \ref{Fig1}, we show the main types of Blockchain forks.
The first category, the temporary forks, are the outcome of a divergence in the consensus process and result in a chain split. Such situations are possible when: \\
 \begin{tabular}{@{$\bullet$ }ll}
& two blocks are discovered at the same time by two different miners;\\
&there is an attack at the consensus level (see \cite{Dumas2021});\\
& there is a time lag in the acceptance of the block (resulting in orphaned or uncle blocks).\\
\end{tabular} \\ 

Why are these forks temporary? Simply because the community will follow the longest chain (considered valid by the majority) while the other one will be abandoned and discontinued. Once the chain split ceases, the consensus process will be unique, and there is no more fork \citep{Bowden2021, Investerest.com2019}.\\

\begin{figure}[h!]
\caption{\label{soft} \textbf{Blockchain Soft Fork} \\ \footnotesize{\textit{Description of a soft fork. Source: adapted from \cite{Bitcoin-Central.com2018}}}}
\centering
\includegraphics[scale=0.68]{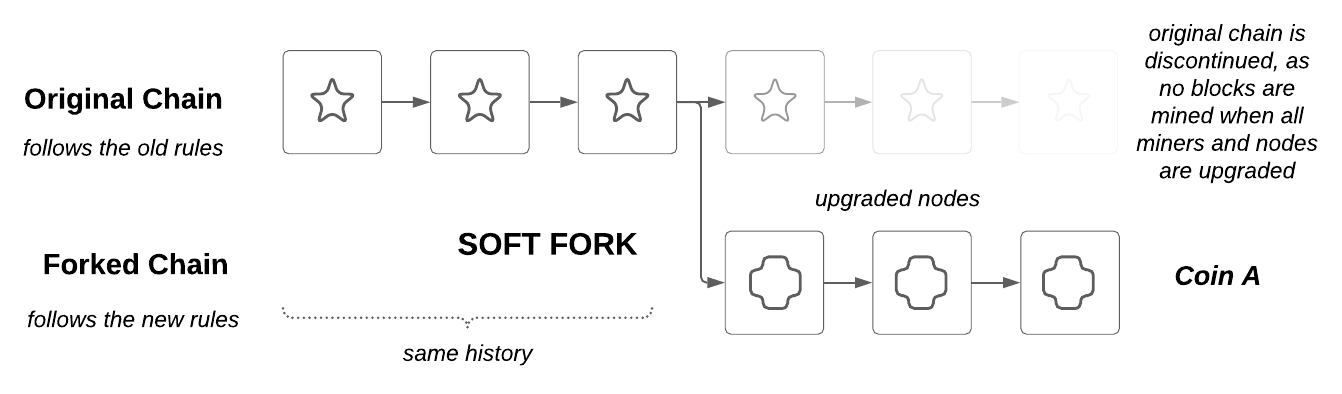}
\end{figure}

On the other hand, permanent forks are due to a change made in the underlying rules of the protocol. These events are planned and pre-announced and sometimes result in a chain split. Considering a software needs, there are situations when it performs upgrading or updating changes. In the case of Blockchain, upgrades are necessary changes to bring an improved and more secure version of the consensus algorithm \citep{Lin2017, Ghosh2020}. These modifications are made in such a way that blocks using the old software will continue to recognize the ones using the new version (it is backward-compatible) and thus resulting in what is called a soft fork \citep{Zhang2017}. For the implementation, the soft fork needs only a majority of participants (51\% within the network) to perform the upgrade. Once this happens, the blocks following the new version of the software will be considered the 'true' ones (therefore, no chain split) \citep{Investerest.com2019, Perez2019}. For better understanding, a visual representation of a soft fork is shown in figure \ref{soft}.\\

\begin{figure}[h!]
\caption{\label{hard} \textbf{Blockchain Hard Fork} \\ \footnotesize{\textit{Description of what is a hard fork. Source: adapted from \cite{Bitcoin-Central.com2018}}}}
\centering
\includegraphics[scale=0.68]{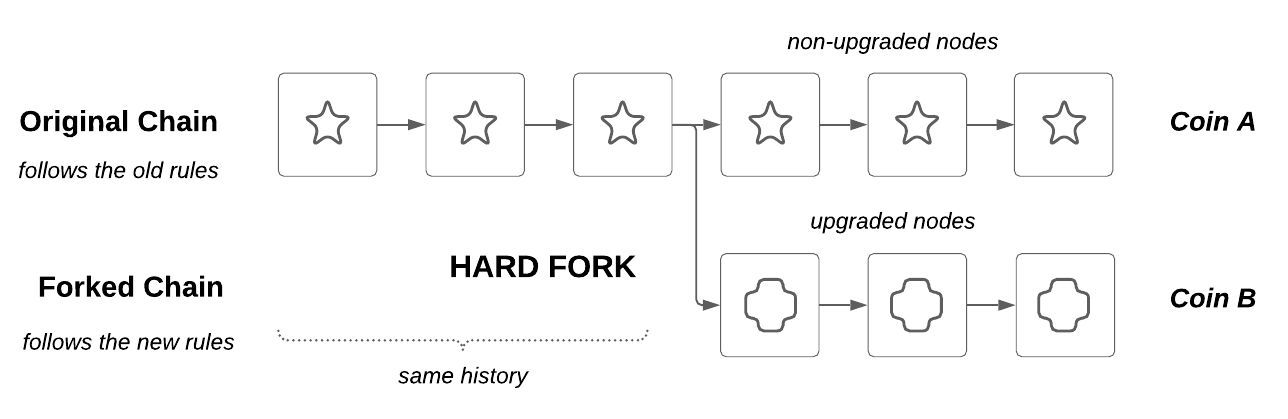}
\end{figure}

Hard forks occur when the consensus algorithm suffers important code modifications (usually for security reasons or to add new functionalities). They can lead to radical protocol changes and a different structure for the Blockchain. Hard forks modifications are not backward-compatible, meaning that the old software is totally distinct from the new one and therefore incompatible \citep{Ghosh2020}. For a successful implementation, hard forks require the contribution of a large subset of participants. In this case, both the new and old software can continue to exist and develop as long as they have enough participants to support them. Here, we are in a scenario where the hard fork generates a chain split and creates a new coin (based on the new Blockchain) \citep{Lin2017}. This scenario is illustrated in figure \ref{hard}. An important mention here is that who owns the \textit{original} coin at the moment of the forking event will receive an equivalent amount of the newly created one.
Now, imagine a scenario when the new software is supported by most of the participants, while the old version by not enough; in this case, the new software will develop as the true chain, while the old version will discontinue as not having enough supporters \citep{Bitcoingold.org2018}. From a technical point of view, this scenario looks similar to figure \ref{soft}, with the mention that the upgraded nodes are not backward-compatible. \\

Blockchain forks, for the most part, occur in a non-random manner. These events are usually planned and discussed within the related cryptocurrency community, such as everyone involved knows what kind of changes must be implemented \citep{Yiu2021}. If looking for possible triggers, we know that the continuous need for improvement as the security and (technological) performance requirements are among the most common reasons behind a permanent fork. From a technical perspective, soft forks and hard forks exhibit notable similarities. However, the first ones represent more a 'cosmetic change', a \textit{slight} and backward-compatible modification in the protocol rules, without affecting the Blockchain structure \citep{Perez2019}. On the other hand, hard forks are more complex changes and require tampering with the Blockchain structure. The complexity of the code modifications can be explained by their needs: to fix bugs, undo illegal transactions (e.g., the DAO attack), increase the throughput, etc. Sometimes, hard forks are considered a solution to disagreements within the community (e.g., Bitcoin Cash vs. Bitcoin SV). Disputes split the participants into different groups, each supporting its own idea of Blockchain development. In these cases, the considered solution is a hard fork that splits the chain and creates a new Blockchain and a new coin. This will allow everyone to follow their ideas and develop the Blockchain independently, as long as there are enough supporters to maintain it \citep{Bitcoingold.org2018, Investerest.com2019}. A detailed list of bitcoin's fork events can be consulted in \cite{BitMEX2017}.\\

In conducting this research, we focus on bitcoin's forking events, them being hard or soft forks.

\subsection{The current state of research}
Despite recent efforts to enrich the literature on cryptocurrencies, we observe that the existing research does not seem to propose enough answers given the market needs. In particular, we mention the relatively scarce work on Blockchain forks. Starting from 2014\footnote{2014 is the year when Ethereum and smart contracts (Blockchain second generation) were created.} and at a faster pace since the Bitcoin Bubble (2017 - 2018), cryptocurrencies are gaining significant attention, provoking an explosion in Blockchain research. Up to now academics have focused on the Bitcoin Bubble \citep{Enoksen2020, Chaim2019}; ICOs \citep{Chohan2019, Chen2020, Adhami2018}; cryptocurrencies' nature \citep{White2016, Nadler2020, Liu2021, Ankenbrand2018, Tan2020}; their volatility \citep{Telli2020, Garcia-Monleon2021, Fakhfekh2020, Kristoufek2019}; and Blockchain attacks \citep{Gramoli2020, Caporale2021}. From the existing literature, we observe that Blockchain forks are mostly treated as either a technological challenge \citep{Vishwanathan2017,  Islam2019, Chen2020, Zamyatin2019, Neudecker2019, Nyman2012, Zhang2017} or a compliance one \citep{Button2019, Xu2019, Webb2018, Schar2020}. In a similar vein, \cite{Button2019} is tackling the effect of hard forks on the crypto holders, \cite{Biais2019a} discuss the miners' vested interests, \cite{Evans2018} shows how the forks' network evolves in time, who are the supporters, and for which reasons they contribute to the network. \cite{Kiffer2017} explores the consequences of a fork on the network, \cite{Azouvi2019} shows that there is little intersection between the communities of the parent coin vs. the forks, and finally, both \cite{Bowden2021}, and \cite{Hotovec2019} show that forks can offer new investment opportunities. More recent research, such as \citep{Bazan-Palomino2021}, compares bitcoin to some of its forks (Litecoin, Bitcoin Cash, Bitcoin Gold, Bitcoin Diamond, Bitcoin Atom, Bitcoin Private, and Bitcoin SV) and concludes that the correlation between bitcoin and the forks is volatility-dependent and that two months after their issuance, the forks contribute strongly to the market volatility.\\

After reviewing the existing literature on Blockchain forks, we have observed that there is little financial research on this topic. Our work aims to fill this gap, and therefore we propose a first assessment of the forking events' impact on the financial characteristics of a cryptocurrency. In this study, we answer the following research question: \textit{How do bitcoin's financial characteristics react to forking events?}. 
Given that forking events are typically driven by technological innovation, we anticipate that these events will be perceived as \textit{good news} and will positively affect bitcoin's performance. 
At the same time, we can look at hard fork events and compare them with their analogous events from the financial market, the spinoff. A spinoff refers to the process in which a company separates a portion of its operations into a new entity and distributes shares of that entity to its existing shareholders. Hard forks can result in chain splits, which lead to the creation of a new Blockchain and a new cryptocurrency. In the stock market, spinoffs enhance investors' wealth, and after these events, usually follows a period of positive abnormal returns \citep{miles1983}. Similarly, in the context of cryptocurrency, investors are financially compensated during a forking event by receiving a proportional amount of the newly created coin based on their holdings of the parent coin. Based on the idea that spinoff events are followed by positive returns \citep{miles1983}, we expect that forks will positively affect bitcoin's returns and volatility. \\

Our work is distinguishable from previous literature in the way that we are the first ones to study the forking effect (the financial impact suffered by a cryptocurrency as a response to forking events). This paper contributes to the understanding of Blockchain forks from both technological and financial points of view.

\section{Data \& Methodology}
In this section, we present the data and methods used to perform this research.

\subsection{Data collection}
This paper studies the forking effect for the bitcoin forks. The choice was mainly made based on the availability of data. Bitcoin is the most known cryptocurrency and the most forked chain. Considering these, any data concerning bitcoin's fork was relatively easy to access. We obtained the bitcoin prices from \href{https://coinmarketcap.com/}{CoinMarketCap.com}, and the CRIX data from \href{https://www.royalton-crix.com/}{Royalton-crix.com} (for data before March 2018) and \href{https://www.spglobal.com/spdji/en/custom-indices/royalton-partners-ag-rpag/royalton-crix-crypto-index/#overview}{spglobal.com} (for data from March 2018 onwards). To gather the necessary information on forked coins, including their names, tickers, and fork dates, we referred to multiple websites, which are listed in the Appendix Section, Table \ref{tab:source}.\\

In this study, our objective is to examine the impact of forking events on bitcoin's returns and volatility. To achieve this, we collected the closing price data for bitcoin/USD from January 1, 2015, to January 1, 2020. We identified a total of 93 forked coins, but due to data availability and reliability concerns, we were able to use only 85 of them\footnote{The sample structure can be consulted in the appendix section, Table \ref{source_part1}.}. To ensure the robustness of our analysis, we excluded early forks (e.g., Litecoin, DigiByte, Dash, etc.) that occurred before 2014, as the financial data during that period was limited and potentially manipulated \citep{LitecoinDeveloper2019}. This decision was supported by the fact that trading data for the crypto-market in the early years was deemed unreliable and subsequently removed from most databases \citep{CoinDesk.com2014, Hileman2013, Partz2018}. Additionally, creating our own crypto-market index to cover the early years was not practical due to the same data issues. As a result, we opted to utilize the CRIX index as a reliable and widely used benchmark for our analysis.\\

\begin{figure}[h!]
\caption {\label{price}{ \textbf{Bitcoin price and forks' dates} \\ \footnotesize{\textit{Chart of the price of bitcoin in US Dollars (BTC/USD) from 01-01-2015 to 01-01-2020. Each fork is represented by a vertical red-dotted line.}}} }
\centering
\includegraphics[scale=0.5]{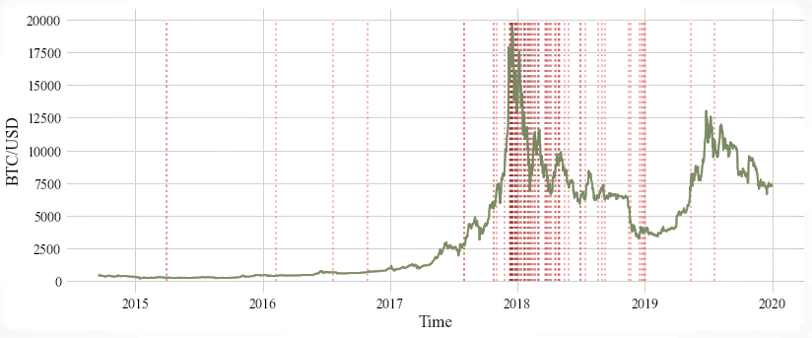}
\end{figure}

\begin{table}[!ht]
\centering
\caption {\label{tab:descriptive}{ \textbf{Descriptive Statistics for BTC and CRIX} \\ \footnotesize{\textit{Descriptive statistics for daily returns of BTC and CRIX from 2014 to 2020. The Jarque-Bera statistics is provided with its corresponding p-value. }}} }
\resizebox{0.45\textwidth}{!}{
\begin{tabular}{lrr}
\toprule
{} &          BTC &         CRIX \\
\midrule
Nb of Obs       &  2297 &  2297 \\
Mean        &     0.002555 &     0.002746 \\
Std Dev         &     0.038435 &     0.038230 \\
Mininum         &    -0.371695 &    -0.360228 \\
Maximum       &     0.252472 &     0.219622 \\
Skewness       &    -0.213003 &    -0.476477 \\
Kurtosis       &     8.587356 &     8.023814 \\
Jarque\_Bera &  7040.180636 &  6220.596245 \\
P-value      &     0.000000 &     0.000000 \\
\bottomrule
\end{tabular}
}
\end{table}

\subsection{Research methodology}
In financial markets, it is commonly assumed that stocks have a fundamental value that represents their actual intrinsic worth. However, due to factors like noise, information asymmetry, temporary illiquidity, and exogenous shocks, stock prices deviate from their fundamental value. Pricing cryptocurrencies is particularly challenging because of their abnormal volatility, which is heavily influenced by investor behavior \citep{Aste2019}. We propose that the fundamental value of cryptocurrencies could be the perceived value of the underlying technology, while fluctuations around this value could stem from disagreements or agreements about the technology's worth. With this perspective, we are interested in exploring the financial implications of fork events in the crypto-market. \\

Forking events in the crypto-market often exhibit a clustering pattern, as shown in Figure \ref{price}. The majority of forks in our sample occur in groups, with 37 forks being followed by another fork in the subsequent days. This clustering characteristic poses a challenge for employing the event study methodology introduced by \cite{MacKinlay1997}. The overlapping windows created by the consecutive forks make it difficult to isolate the impact of a single fork from the effects of the forks that occurred on the previous or following days. To tackle the intricacies of our dataset, we have opted for a methodology inspired by the research conducted by \cite{Grobys2021}. This approach enables us to effectively address the complexities and unique characteristics of the data we are analyzing. \cite{Grobys2021} studied the impact of hacking events on bitcoin's volatility using a variant of the Generalised Auto Regressive Conditional Heteroskedasticity (GARCH) model, namely EGARCH.  With this analysis, we examine the influence of forking events on both the returns and volatility of bitcoin. The EGARCH model can be modified to include the effect of a dummy variable, in our case forking events, on either the returns or the volatility of a cryptocurrency.  
Contrary to the typical positive coefficient observed in stock markets, \cite{baur2018} demonstrate that negative shocks have a smaller impact on volatility compared to positive shocks in the crypto-market. Given that forking events are typically driven by technological innovation, we anticipate that these events will be perceived as \textit{good news} and will positively affect bitcoin's performance. To capture the asymmetric volatility, we employ the exponential GARCH (EGARCH) model, which was originally introduced by \cite{Nelson1991}. This modeling approach enables us to examine the specific impact of forks on bitcoin's performance.

\subsection{Returns' reaction to a forking event }

We estimate a modified version of the EGARCH model similar to \cite{Grobys2021}. In this model, we assume that asset returns follow a process (as shown in the first equation of the set of Equations \ref{eq:egarch_mean}). The average return is denoted by $\mu$ and is influenced by technological events such as forks ($D(t)$) and the overall market conditions ($R_{CRIX}(t)$). Additionally, returns are subject to shocks that depend on the conditional volatility of asset returns ($\sigma(t)$). The volatility process has an unconditional mean of $\omega$ and exhibits asymmetric responses to shocks, meaning that a large positive return does not have the same impact on volatility as a large negative return. Furthermore, the volatility process is influenced by the previous day's volatility, leading to volatility clustering. It is common to consider fat-tail distributions for the shock on returns; therefore, we adopt a student distribution with degrees of freedom equal to 5\footnote{\cite{Grobys2021} is studying the impact of bitcoin hacking events on bitcoin data. Considering the similarities in the data between our study and \citep{Grobys2021}, we decided to use the same value for the degrees of freedom.}, following the approach of \cite{Grobys2021}.

\begin{equation}
\label{eq:egarch_mean}
\begin{aligned}
    & R(t) = \mu + \mathbf{\delta_{fork\_mean} D(t)} + \delta_{CRIX} R_{CRIX}(t) + \varepsilon(t)\\
    & \varepsilon(t) = \sigma(t) z(t) \\
    & Where :\ z(t) | \Omega_{t-1} \sim t(\nu) \\
    & ln(\sigma^2(t)) = \omega + \alpha \Bigl( |z(t-1)| - \mathbf{E}[|z(t-1)|] \Bigl) +  \gamma z(t-1) + \beta ln(\sigma^2(t-1)) \\
\end{aligned}
\end{equation}

In the Equation set \ref{eq:egarch_mean}, the $R(t)$ is the vector of cryptocurrency returns (BTC), $\mu$ is the expected return, $D(t)$ is the dummy variable for fork events; therefore $D(t)$  takes the value 1 for the days when we have a forking event and takes the value 0 when there are no events. $R_{CRIX}(t)$ is the vector of CRIX returns (our market index), $\sigma^2$ is the conditional variance, $z(t)$ is a Student innovation process with $\nu$ degrees of freedom, $\Omega_{t-1}$ is the information set at $t-1$, and $ [\mu, \delta_{fork\_mean}, \delta_{CRIX}, \omega, \alpha, \gamma, \beta]$ is the vector of parameters to be estimated via Quasi Maximum Likelihood Estimation (QMLE)\footnote{The QLME provides robust standard errors for the coefficients of the model as it does not require distributional assumptions to hold.}. Similarly to \cite{Grobys2021}, we set the degrees of freedom ($\nu$) to be 5. \\

In conducting this analysis, we differ from \cite{Grobys2021} by running our model twice: firstly, similar to  \cite{Grobys2021} without considering the market factor, and secondly, incorporating the market dynamics by considering the CRIX index. The motivation behind this choice is as follows. As depicted in Figure \ref{price}, the majority of forking events occur during the Bitcoin Bubble period. Given the significance of market risk premium as a predictor of cryptocurrency returns \citep{dunbar}, we deemed it relevant to explore any potential differences between these two models. Furthermore, we differ from the traditional EGARCH model, which was originally developed by \cite{Nelson1991}, by choosing the innovation process ($z(t)$) to follow a Student distribution. This choice is motivated by the observation that cryptocurrency returns, particularly bitcoin returns in our sample, exhibit high kurtosis. The excess kurtosis of our bitcoin returns is measured at 5.32. By employing the Student distribution, we can account for the presence of fat-tails in the return distribution \citep{Grobys2021}, capturing the occurrence of extreme events more accurately.\\

Drawing on the theoretical framework outlined earlier, we formulate and examine the following hypotheses:

\begin{equation}
\label{eq:hypo_mean}
\begin{aligned}
    & H0: \delta_{fork\_mean} = 0 \\
    & H1: \delta_{fork\_mean} \neq 0 \\
\end{aligned}
\end{equation}

Where $ \delta_{fork\_mean} = 0$ means that forking events do not impact BTC returns, and $\delta_{fork\_mean} \neq 0$ means that forks impact BTC returns. 

\subsection{Volatility's reaction to a forking event}

In a manner analogous to our investigation of the effect of forks on returns, we employ a similar approach to examine the impact of fork events on the volatility of the parent coin. To accomplish this, we utilize a modified version of the EGARCH model similar to \citep{Grobys2021}, which considers the presence of asymmetrical shocks in the conditional variance. We modify the eGARCH model to suit our analysis as follows:

\begin{equation}
\label{eq:egarch_var}
\begin{aligned}
    & R(t) = \mu  + \delta_{CRIX} R_{CRIX}(t) + \varepsilon(t)\\
    & \varepsilon(t) = \sigma(t) z(t) \\
    & Where :\ z(t) | \Omega_{t-1} \sim t(\nu) \\
    & ln(\sigma^2(t)) = \omega + \alpha \Bigl( |z(t-1)| - \mathbf{E}[|z(t-1)|] \Bigl) +  \gamma z(t-1) + \beta ln(\sigma^2(t-1)) + \mathbf{\delta_{fork\_variance} D(t)} \\
\end{aligned}
\end{equation}

The description of the variables and parameters remains the same as for the equation \ref{eq:egarch_mean}, except for the vector of parameters $[\mu, \delta_{CRIX}, \omega, \alpha, \gamma, \beta, \delta_{fork\_variance}]$. We conduct two tests of this model: one includes the market index $R_{CRIX}(t)$, while the other excludes it. By employing this methodology, we are able to examine how market dynamics influence the impact of a fork on the volatility of the parent coin. \\

To evaluate the effect of forking events on volatility, we put forward the following hypotheses for testing: 

\begin{equation}
\label{eq:hypo_variance}
\begin{aligned}
    & H0: \delta_{fork\_variance} = 0 \\
    & H2: \delta_{fork\_variance} \neq 0 \\
\end{aligned}
\end{equation}

Where $ \delta_{fork\_variance} = 0$ means that forking events do not impact BTC volatility, and $\delta_{fork\_variance} \neq 0$ means that forks impact BTC volatility. 

\section{Results}
\label{result}

The discussion of the results will be separated into two parts. One detailing the impact that forking events have on bitcoin's returns and the second about the impact on volatility. 

\subsection{Impact on Returns}

We initially tested our model (see Equation \ref{eq:egarch_mean}) by excluding the returns of the CRIX. We wanted to see whether the forking events have any effect on the overall return of the parent coin (bitcoin). The model was calibrated to incorporate fat tails, addressing a common issue noted by \cite{taleb2020} on inference in the presence of fat-tailed distributions. The degrees of freedom used for the Student innovation process is $\nu=5$ as the original methodology of \cite{Grobys2021}. The results are shown in Table \ref{table:EGARCH_mean}. We can observe that the estimated coefficient $\gamma$ is positive and significant at any conventional significance level, indicating a strong positive asymmetric response in bitcoin's volatility. Similarly, the estimated coefficient $\beta$ is  highly significant and close to 1, indicating the presence of volatility clustering. These findings align with the results reported by \cite{Grobys2021} and provide further support for the validity of our model. Our results show as well the fact that there is no significant response in the bitcoin's returns in reaction to the forking events (the $\delta_{fork\_mean}$ is close to zero and not significant). 
Since we don't find any significant impact on bitcoin returns, we think that this result suggests the fact that investors do not perceive forks as either positive or negative news. We may think that investors are in fact insensitive to the forking events.

\begin{table}[ht!]
\centering
\caption{\label{table:EGARCH_mean} \textbf{Coefficients estimate of the EGARCH(1,1) - Returns} \\ \footnotesize{\textit{The table below shows the estimate values of the coefficients of the EGARCH(1,1) model used to evaluate the impact of forks on the parent coin's returns.}}}
\resizebox{0.75\textwidth}{!}{
\begin{tabular}{l|llll}
\textit{\textbf{Parameter}}             & \textit{\textbf{Estimate}} & \textit{\textbf{Std. Error (in \%)}} & \textit{\textbf{t value}} & \textit{\textbf{p value}} \\ \hline
\textbf{$\mu$}                          & \textbf{0.0015}            & \textbf{0.049}                       & \textbf{2.956}            & \textbf{0.003}            \\
\textbf{${\delta_{fork\_mean}}$} & 0.0039                     & 0.541                                & 0.722                     & 0.471                     \\
\textbf{$\omega$}                       & -0.1298                    & 8.503                                & -1.526                    & 0.127                     \\
\textbf{$\alpha$}                       & 0.0206                     & 1.708                                & 1.203                     & 0.229                     \\
\textbf{$\beta$}                        & \textbf{0.9826}            & \textbf{1.231}                       & \textbf{79.776}           & \textbf{0.000}            \\
\textbf{$\gamma$}                       & \textbf{0.2337}            & \textbf{5.337}                       & \textbf{4.379}            & \textbf{0.000}           
\end{tabular}
}
\end{table}

 In Table \ref{table:EGARCH_mean_CRIX}, we estimate our model taking into account the market dynamics and find that the asymmetric response in volatility is stronger than before ($\gamma=0.2446^{***}$) and the volatility clustering is still present ($\beta=0.9861^{***}$). At the same time, we see that the $\delta_{CRIX}$ is close to 1 and strongly significant. Therefore, in such a setting, this model studies the impact of forks on the excess returns of bitcoin with the market. If $\delta_{CRIX} \approx 1$, then we find the following:  $R(t) = \mu + \delta_{fork\_mean} D(t) + \delta_{CRIX} R_{CRIX}(t) + \varepsilon(t)$ is equivalent to $R(t) - R_{CRIX}(t) = \mu + \delta_{fork\_mean} D(t) + \varepsilon(t)$. Despite the decrease in the standard error of $\delta_{fork\_mean}$, it is not statistically significant, indicating that forking events do not have an immediate impact on the actual returns of the parent coin. It is worth noting that there might be a delayed effect of forking events on returns, which warrants further investigation. Additional research is necessary to delve deeper into this aspect and explore the potential lagged effects of forking events on returns. \\

Based on our analysis, we fail to reject the null hypothesis stated in Equation \ref{eq:hypo_mean}, as indicated by the p-value of $17.6\%$. This result suggests that there is insufficient evidence to conclude that forks have a significant impact on the average returns of bitcoin when they occur. \\

\begin{table}[ht!]
\centering
\caption{\label{table:EGARCH_mean_CRIX} \textbf{Coefficients estimate of the EGARCH(1,1) with CRIX - Returns} \\ \footnotesize{\textit{The table below shows the estimate values of the coefficients of the EGARCH(1,1) model used to evaluate the impact of forks on the parent coin's returns taking into account the market dynamics.}}}
\resizebox{0.77\textwidth}{!}{
\begin{tabular}{l|llll}
\textit{\textbf{Parameter}}             & \textit{\textbf{Estimate}} & \textit{\textbf{Std. Error (in \%)}} & \textit{\textbf{t value}} & \textit{\textbf{p value}} \\ \hline
\textbf{$\mu$}                          & \textbf{0.0003}            & \textbf{0.016}                       & \textbf{2.041}            & \textbf{0.041}            \\
\textbf{$ \mathbf{\delta_{CRIX}}$}                & \textbf{0.964}             & \textbf{1.3718}                      & \textbf{70.253}           & \textbf{0.000}            \\
\textbf{${\delta_{fork\_mean}}$} & -0.0035                    & 0.225                                & -1.354                    & 0.176                     \\
\textbf{$\omega$}                       & \textbf{-0.1153}           & \textbf{1.139}                       & \textbf{10.124}           & \textbf{0.000}            \\
\textbf{$\alpha$}                       & \textbf{0.0755}            & \textbf{2.586}                       & \textbf{2.918}            & \textbf{0.004}            \\
\textbf{$\beta$}                        & \textbf{0.9861}            & \textbf{0.127}                       & \textbf{775.702}          & \textbf{0.000}            \\
\textbf{$\gamma$}                       & \textbf{0.2446}            & \textbf{2.795}                       & \textbf{8.754}            & \textbf{0.000}           
\end{tabular}
}
\end{table}

\subsection{Impact on Volatility}

In our analysis, we have investigated the effect of forking events on the volatility of the parent coin by utilizing the model presented in Equation \ref{eq:egarch_var}. Our findings reveal the presence of asymmetry in volatility ($\gamma=0.2330^{***}$) and volatility clustering ($\beta=0.9768^{***}$), which aligns with previous research \citep{baur2018, katsiampa2019, cheikh2020, Grobys2021}. However, surprisingly, we observe that the scale of the innovation, as indicated by $\alpha$, does not have a significant impact on volatility. Interestingly, we find that forking events do have an immediate impact on the volatility of bitcoin, as evidenced by the significant coefficient $\delta_{fork\_variance}=0.1508^{**}$. This suggests that forks lead to an increase in volatility on the day they occur. An interesting idea would be to explore the potential longer-term effects of forks on bitcoin's volatility, considering the overlapping nature of these events. This could involve estimating the impact of a fork on bitcoin's volatility in the days following the event.

\begin{table}[ht!]
\centering
\caption{\label{table:EGARCH_Volatility} \textbf{Coefficients estimate of the EGARCH(1,1) - Volatility} \\ \footnotesize{\textit{The table below shows the estimate values of the coefficients of the EGARCH(1,1) model used to evaluate the impact of forks on the parent coin's volatility.}}}
\resizebox{0.77\textwidth}{!}{
\begin{tabular}{l|llll}
\textit{\textbf{Parameter}}                 & \textit{\textbf{Estimate}} & \textit{\textbf{Std. Error (in \%)}} & \textit{\textbf{t value}} & \textit{\textbf{p value}} \\ \hline
\textbf{$\mu$}                              & \textbf{0.0015}            & \textbf{0.042}                       & \textbf{3.615}            & \textbf{0.000}            \\
\textbf{$\omega$}                           & \textbf{-0.1751}           & \textbf{5.021}                       & \textbf{-3.487}           & \textbf{0.000}            \\
\textbf{$\alpha$}                           & 0.0227                     & 1.592                                & 1.426                     & 0.154                     \\
\textbf{$\beta$}                            & \textbf{0.9768}            & \textbf{0.719}                       & \textbf{135.816}          & \textbf{0.000}            \\
\textbf{$\gamma$}                           & \textbf{0.2330}            & \textbf{3.410}                       & \textbf{6.833}            & \textbf{0.000}            \\
 \textbf{$\mathbf{\delta_{fork\_variance}}$} & \textbf{0.1508}            & \textbf{5.955}                       & \textbf{2.533}            & \textbf{0.011}           
\end{tabular}
}
\end{table}

In Figure \ref{price}, we show that most of the forking events happen during the Bitcoin Bubble. Consequently, we can expect that the volatility during this period will be higher than usual and might bias our results. To address this issue, we incorporate market dynamics in our model and obtain the results displayed in Table \ref{table:EGARCH_Volatility_CRIX}. Our findings confirm again the presence of asymmetric response and volatility clustering ($\gamma=0.2408^{***}$ and $\beta=0.9838^{***}$), which is in line with the existing literature \citep{baur2018, katsiampa2019, cheikh2020, Grobys2021}. Furthermore, we observe that the effect the forking event has on the volatility is strengthened. The $\delta_{fork\_variance}$ is now $0.2005^{***}$. This means that the forking events' impact on volatility is stronger when we consider market dynamics. The residual volatility (the one that is not due to the market) still peaks on the day of the forking event.

\begin{table}[ht!]
\centering
\caption{\label{table:EGARCH_Volatility_CRIX} \textbf{Coefficients estimate of the EGARCH(1,1) with CRIX - Volatility} \\ \footnotesize{\textit{The table below shows the estimate values of the coefficients of the EGARCH(1,1) model used to evaluate the impact of forks on the parent coin's volatility taking into account the market dynamics.}}}
\resizebox{0.75\textwidth}{!}{
\begin{tabular}{l|llll}
\textit{\textbf{Parameter}}        & \textit{\textbf{Estimate}} & \textit{\textbf{Std. Error (in \%)}} & \textit{\textbf{t value}} & \textit{\textbf{p value}} \\ \hline
\textbf{$\mu$}                     & 0.0004                     & 0.016                                & 1.924                     & 0.054                     \\
\textbf{$\mathbf{\delta_{CRIX}}$}  & \textbf{0.9651}            & \textbf{1.383}                       & \textbf{69.815}           & \textbf{0.000}            \\
\textbf{$\omega$}                  & \textbf{-0.1399}           & \textbf{0.913}                       & \textbf{-15.314}          & \textbf{0.000}            \\
\textbf{$\alpha$}                  & \textbf{0.0713}            & \textbf{2.211}                       & \textbf{3.226}            & \textbf{0.001}            \\
\textbf{$\beta$}                   & \textbf{0.9838}            & \textbf{0.100}                       & \textbf{1068.673}         & \textbf{0.000}            \\
\textbf{$\gamma$}                  & \textbf{0.2408}            & \textbf{2.598}                       & \textbf{9.269}            & \textbf{0.000}            \\
\textbf{$\mathbf{\delta_{fork\_variance}}$} & \textbf{0.2005}            & \textbf{6.167}                       & \textbf{3.252}            & \textbf{0.001}           
\end{tabular}
}
\end{table}

Overall, our results show that the uncertainty increases when a forking event occurs. We also estimated our model with another variable that counts how many forks occurred each day, and we found that the relationship is not more significant than the result displayed in Table \ref{table:EGARCH_Volatility_CRIX}. This shows that the uncertainty coming from a fork does not depend on how many forks are actually taking place\footnote{For more details on why the model with the dummy variable dominates, please refer to the Appendix, Table \ref{table:Model_Choice}.}. A possible interpretation could be that investors make short-term choices on which Blockchain to follow. Regardless, our results confirm that, when studying the volatility of cryptocurrencies, one should use a model that accounts for asymmetric responses in volatility and, furthermore, should consider the market dynamics. Regarding our initial hypothesis (see Equation \ref{eq:hypo_variance}), we can reject $H0$ and hence validate that there is evidence to believe that forking events positively impact the volatility of the parent coin at any level of significance.

\section{Additional tests and results}
 In order to provide more insights about our results, we investigate whether volatility tends to be higher when multiple forks occur on the same day, whether the increase of volatility compounds if forks occur on consecutive days, and whether our findings are robust when only accounting for hard forks. 

\subsection{The effect of multiple forking events occurring on the same day}

Given the fact that when a forking event takes place, the volatility in that specific day increases, we could expect the volatility to increase even more if more than one fork occurs simultaneously. The rationale is straightforward: if one fork causes uncertainty in the market, then multiple forking events happening on the same day should translate into more uncertainty. To test this hypothesis, we construct multiple two samples Welch t-tests (see Table \ref{tab:differences_in_number}). In order to examine potential differences in average volatility, we compared four scenarios: (1) forks that occur alone (One fork) with events that occur in pairs (two forks), (2) one fork with events that occur in triplets (three forks), (3) two forks with three forks, and (4) one fork with multiple forks. Results are reported in Table \ref{tab:differences_in_number}. 

\begin{table}[h!]
\centering
\caption {\label{tab:differences_in_number}{ \textbf{Variations in volatility based on the number of forking events happening within a day} \\ \footnotesize{\textit{Here we have used the Welch two samples t-tests for different cases. For example, test 1 shows differences between days when only one fork occurs and days when two forks occur. The column difference shows the differences in average volatility between variable 1 and variable 2. The t-tests are two-sided.}}} }
\resizebox{0.78\textwidth}{!}{
\begin{tabular}{c|llrrr}
\textit{\textbf{Test \#}} & \textit{\textbf{Variable 1}} & \textit{\textbf{Variable 2}} & \textit{\textbf{Difference}} & \textit{\textbf{t value}} & \textit{\textbf{p value}} \\ \hline
\textbf{1}                & One fork                     & Two forks                    & -0.003                       & -0.2607                   & 0.7983                    \\
\textbf{2}                & One fork                     & Three forks                  & 0.018                        & 1.459                     & 0.2612                    \\
\textbf{3}                & Two forks                    & Three forks                  & 0.020                        & 1.378                     & 0.2285                    \\
\textbf{4}                & One fork                     & Multiple forks               & 0.002                        & 0.221                     & 0.8265                   
\end{tabular}
}
\end{table}

The results from our test conclude that the number of forks taking place on the same day does not change the impact on bitcoin's volatility. So, we can conclude that the uncertainty created by a fork does not compound with other forks taking place on the same day. 

\subsection{Fork clusters \& delayed effects }

The previous section shows that multiple forks occurring on the same day do not compounds the impact one has on the bitcoin's volatility. Here, we examine the persistence of a fork's impact on volatility in the days following the event, as well as the cumulative effect of multiple forks occurring consecutively. We categorize forks into two types: those that are not followed by any subsequent events (no subsequent forks) and those that are followed by other events (subsequent forks). We conduct t-tests to compare the average volatility on the days immediately following the first fork event (see Table \ref{tab:delayed_effect}). 

\begin{table}[!h]
\caption {\label{tab:delayed_effect}{ \textbf{Compound and delay effects when multiple forking events occur} \\ \footnotesize{\textit{In this analysis, we examine the average volatility changes following a fork event under two scenarios. The first scenario considers the case where no subsequent fork occurs after the initial event. The second scenario considers the case where a subsequent fork occurs on the 1st, 2nd, or 3rd day following the initial fork. We conduct two-sided Welch t-tests and report the corresponding t-values and p-values to assess the significance of the observed changes in average volatility.}}} }
\resizebox{1\textwidth}{!}{
\begin{tabular}{l|rrrr|rrrr}
\textit{\textbf{}} & \multicolumn{4}{c|}{\textbf{No subsequent forks}}                                                                & \multicolumn{4}{c}{\textbf{Subsequent forks}}                                                                    \\ \hline
                   & \textit{\textbf{Av. vol.}} & \textit{\textbf{Std error}} & \textit{\textbf{t value}} & \textit{\textbf{p value}} & \textit{\textbf{Av. vol.}} & \textit{\textbf{Std error}} & \textit{\textbf{t value}} & \textit{\textbf{p value}} \\ \hline
\textbf{t}         & 0.0589                     & 0.0025                      &                           &                           & 0.0589                     & 0.0025                      &                           &                           \\
\textbf{t+1}       & 0.0599                     & 0.0021                      & -0.1777                   & 0.8596                    & 0.0618                     & 0.0012                      & -0.4691                   & 0.6413                    \\
\textbf{t+2}       & 0.0592                     & 0.0022                      & -0.0659                   & 0.9477                    & 0.0618                     & 0.001                       & -0.5196                   & 0.6058                    \\
\textbf{t+3}       & 0.0585                     & 0.0022                      & 0.0639                    & 0.9493                    & 0.0614                     & 0.0009                      & -0.4576                   & 0.6495                   
\end{tabular}
}
\end{table}

Our results show that after a forking event, volatility tends to stay high for the following three days, regardless of whether other events occur or not.

\subsection{Robustness checks}
To ensure the reliability and validity of our findings, we perform robustness checks by making our tests just on the data from hard forks. In our sample, we accounted for 22 hard forks (data is presented in \ref{tab:hrdf}). We remind the fact that hard forks are those types of event that often lead to a chain split and the creation of a new cryptocurrency. We run the models used before (see Equations \ref{eq:egarch_mean} and \ref{eq:egarch_var}) to study the impact the hard forks have on bitcoin's returns and volatility. For the investigation of the  hard forks' impact on bitcoin's returns, the results are displayed in Table \ref{table:HardFork_EGARCH_mean_CRIX}.  For the investigation on bitcoin's volatility, see Table \ref{table:HardFork_EGARCH_Volatility_CRIX}.

\begin{table}[ht!]
\centering
\caption{\label{table:HardFork_EGARCH_mean_CRIX} \textbf{Coefficients estimate of the EGARCH(1,1) with CRIX - Returns for hard forks} \\ \footnotesize{\textit{The table below shows the estimate values of the coefficients of the EGARCH(1,1) model used to evaluate the impact of hard forks on the parent coin's returns taking into account the market dynamics. The model is run on a sample of 22 hard forks.}}}
\resizebox{0.73\textwidth}{!}{
\begin{tabular}{l|llll}
\textit{\textbf{Parameter}}             & \textit{\textbf{Estimate}} & \textit{\textbf{Std. Error (in \%)}} & \textit{\textbf{t value}} & \textit{\textbf{p value}} \\ \hline
\textbf{$\mu$}                          & \textbf{0.0003}            & \textbf{0.0151}                      & \textbf{2.118}            & \textbf{0.034}            \\
\textbf{$\delta_{CRIX}$}                & \textbf{0.964}             & \textbf{1.0201}                      & \textbf{94.4681}          & \textbf{0.000}            \\
\textbf{$\mathbf{\delta_{fork\_mean}}$} & -0.0033                    & 0.3147                               & -1.065                    & 0.287                     \\
\textbf{$\omega$}                       & \textbf{-0.1133}           & \textbf{1.2288}                      & \textbf{-9.227}           & \textbf{0.000}            \\
\textbf{$\alpha$}                       & \textbf{0.0750}            & \textbf{1.9244}                      & \textbf{3.898}            & \textbf{0.000}            \\
\textbf{$\beta$}                        & \textbf{0.9863}            & \textbf{0.1463}                      & \textbf{674.093}          & \textbf{0.000}            \\
\textbf{$\gamma$}                       & \textbf{0.2436}            & \textbf{2.1495}                      & \textbf{11.331}           & \textbf{0.000}           
\end{tabular}
}
\end{table}

\begin{table}[ht!]
\centering
\caption{\label{table:HardFork_EGARCH_Volatility_CRIX} \textbf{Coefficients estimate of the EGARCH(1,1) with CRIX - Volatility for hard forks} \\ \footnotesize{\textit{The table below shows the estimate values of the coefficients of the EGARCH(1,1) model used to evaluate the impact of hard forks on the parent coin's volatility taking into account the market dynamics. The model is run on a sample of 22 hard forks.}}}
\resizebox{0.75\textwidth}{!}{
\begin{tabular}{l|llll}
\textit{\textbf{Parameter}}             & \textit{\textbf{Estimate}} & \textit{\textbf{Std. Error (in \%)}} & \textit{\textbf{t value}} & \textit{\textbf{p value}} \\ \hline
\textbf{$\mu$}                          & \textbf{0.0003}            & \textbf{0.0105}                      & \textbf{2.823}            & \textbf{0.004}            \\
\textbf{$\delta_{CRIX}$}                & \textbf{0.9651}            & \textbf{1.0152}                      & \textbf{95.065}           & \textbf{0.000}            \\
\textbf{$\omega$}                       & \textbf{-0.1173}           & \textbf{1.2645}                      & \textbf{-9.274}           & \textbf{0.000}            \\
\textbf{$\alpha$}                       & \textbf{0.0700}            & \textbf{1.9364}                      & \textbf{3.617}            & \textbf{0.000}            \\
\textbf{$\beta$}                        & \textbf{0.9865}            & \textbf{0.1488}                      & \textbf{662.782}          & \textbf{0.000}            \\
\textbf{$\gamma$}                       & \textbf{0.2308}            & \textbf{2.2493}                      & \textbf{10.261}           & \textbf{0.000}            \\
\textbf{$\mathbf{\delta_{fork\_variance}}$} & \textbf{0.3961}        & \textbf{13.9055}                     & \textbf{2.8499}           & \textbf{0.004}           
\end{tabular}
}
\end{table}

As we can observe in Table \ref{table:HardFork_EGARCH_mean_CRIX}, hard fork events do not impact bitcoin's returns, a result in line with our initial investigation from Section \ref{result}. At the same time, when we look at Table \ref{table:HardFork_EGARCH_Volatility_CRIX}, our analysis of the  hard forks' impact on bitcoin's volatility shows highly significant results. Here, the results of our tests are stronger and more significant than our initial findings from Section \ref{result}, which can be justified by the economic implications of hard fork events, notably the creation of new cryptocurrencies. In conclusion, we confirm that our results are robust.

\section{Conclusion}
The crypto-market presents a significant challenge for both finance academics and practitioners, as it defies conventional financial market norms and principles. Several research studies have attempted to establish connections between cryptocurrencies and other types of existing asset classes \citep{White2016, Ankenbrand2018, Nadler2020, Liu2021, Tan2020}, to propose coherent valuation methods adapted to the crypto-market \citep{Pagnotta2022, Cong2021} or to study the chaotic price dynamics of crypto-assets \citep{Sornette2014, Chaim2019, Enoksen2020}. In the end, it seems that the key to understanding this peculiar market lies in our comprehension of the underlying technology, namely Blockchain, and how it impacts different financial variables. To highlight the causal relationship between technological features and financial dynamics, we propose to study an event specific to cryptocurrencies: the forks. In the Blockchain world, a fork represents a modification, a discrepancy, or a breach of its consensus protocol (sometimes resulting in a new coin).\\

The objective of this study is twofold. Firstly, we investigate the impact of forking events on bitcoin returns, expecting a positive effect based on the association with technological innovation and the positive impact observed in spinoff events  \citep{miles1983}. However, our findings indicate that investors are insensitive to forking events, as no significant impact is observed on bitcoin returns. Secondly, we analyze the effect of forking events on bitcoin volatility and find that they lead to increased market uncertainty and higher volatility. The volatility response is even stronger when considering market dynamics.
To complement our initial results, we further investigated the effects of multiple forks occurring on the same day, the compounding volatility impact of consecutive-day forks, and the robustness of our findings when considering only hard forks. Our final results indicate that: (1) the uncertainty generated by a fork does not intensify when multiple forks occur simultaneously; (2) following a forking event, volatility tends to remain elevated for the subsequent three days, irrespective of the occurrence of other events; (3) the robustness checks conducted validate the reliability and validity of our results, as consistent findings were obtained when analyzing all forking events together or exclusively hard forks. \\

This paper contributes to the understanding of Blockchain forks from both technological and financial points of view. Besides the contributions to the crypto-related literature, the results obtained may have important implications for crypto-investors, who need to take into account the effect of technological events to be able to efficiently mitigate the risks from this market. \\

One limitation of this study is its focus exclusively on forking events in the context of bitcoin, which may restrict the generalizability of the findings to other cryptocurrencies. Therefore, as a future path for research, it would be interesting to see how the forking effect impacts other cryptocurrencies that have been forked, such as ether, litecoin, monero, and others. However, constructing such a database for other coins could be challenging, as relevant information concerning altcoins is usually less centralized, and more difficult to verify its authenticity than for bitcoin. Other interesting paths for  research would be to estimate the short-term, lasting effect of a forking event on the parent coin and compare the long-term performance of the forked coins with their parent. \\

\section*{CRediT authorship contribution statement}
\textbf{Florentina Soiman}: Conceptualization, Methodology, Writing - Original draft, Data extraction, and Analysis. 
 \textbf{Mathis Mourey}: Methodology, Writing - Original draft, Data extraction, and Analysis. 
 \textbf{Jean-Guillaume Dumas}: Supervision, Reviewing, and Editing. \textbf{Sonia Jimenez-Garces}: Conceptualization, Supervision, Validation, Reviewing, and Editing.

\section*{Acknowledgments}
Many thanks to the participants of the World Finance Conference from 2021 and the 39 French Finance Association (AFFI) conference from 2023 for their valuable comments and discussions.  


\pagebreak
\bibliography{library}
\bibliographystyle{apalike}

\newpage
\appendix

\setcounter{figure}{0}
\setcounter{table}{0}
\renewcommand{\thetable}{A.\arabic{table}}

\section{Tables}
\label{appendix:Tables}
\begin{table}[ht!]
\centering
\small
\caption{\label{source_part1}\textbf{The list of bitcoin's forks} \\ \footnotesize{\textit{Comprehensive list of all the forks considered in our study. We provide the ticker as well as the date of the forking event.}}}
\begin{minipage}{0.35\textwidth}
\fontsize{11pt}{11pt}\selectfont
\begin{tabular}{lll}
\toprule
\textbf{Fork name} & ticker &  Fork date \\
\midrule
\textbf{Anonymous Bitcoin     } &   ANON & 2018-09-10 \\
\textbf{Big Bitcoin           } &    BBC & 2018-02-12 \\
\textbf{Bitclassic Coin       } &   BICC & 2017-12-12 \\
\textbf{Bitcoin 2             } &   BTC2 & 2018-02-05 \\
\textbf{Bitcoin Air           } &    XAP & 2018-11-22 \\
\textbf{Bitcoin Atom          } &    BCA & 2018-01-24 \\
\textbf{Bitcoin Blvck         } &   BTCV & 2018-02-05 \\
\textbf{Bitcoin Boy           } &    BCB & 2018-01-02 \\
\textbf{Bitcoin Cash          } &    BCH & 2017-08-01 \\
\textbf{Bitcoin SV            } &    BSV & 2017-08-01 \\
\textbf{Bitcoin Cbc           } &   BCBC & 2017-12-11 \\
\textbf{Bitcoin Clashic       } &   BCHC & 2017-08-01 \\
\textbf{Bitcoin Clean         } &    BCL & 2018-04-18 \\
\textbf{Bitcoin Cloud         } &    BCL & 2018-02-20 \\
\textbf{Bitcoin Community     } &   BTSQ & 2018-01-25 \\
\textbf{Bitcoin Coral         } &   BTCO & 2017-10-24 \\
\textbf{Bitcoin Dao           } &    BTD & 2018-06-30 \\
\textbf{Bitcoin Diamond       } &    BCD & 2017-11-24 \\
\textbf{Bitcoin Dollar        } &    BTD & 2018-02-28 \\
\textbf{Bitcoin Eco           } &    BEC & 2018-12-18 \\
\textbf{Bitcoin Faith         } &    BTF & 2017-12-18 \\
\textbf{Bitcoin File          } &   BIFI & 2017-12-27 \\
\textbf{Bitcoin Flash         } &    BTF & 2018-02-06 \\
\textbf{Bitcoin God           } &    GOD & 2017-12-27 \\
\textbf{Bitcoin Gold          } &    BTG & 2017-10-24 \\
\textbf{Bitcoin Holocaust     } &  BTHOL & 2017-12-29 \\
\textbf{Bitcoin Hot           } &    BTH & 2017-12-12 \\
\textbf{Bitcoin Hush          } &   BTCH & 2018-02-01 \\
\textbf{Bitcoin Interest      } &    BCI & 2018-01-20 \\
\textbf{Bitcoin King          } &    BCK & 2017-12-18 \\
\textbf{Bitcoin Lambo         } &    BTL & 2018-03-27 \\
\textbf{Bitcoin Lightning     } &    BLG & 2017-12-10 \\
\textbf{Bitcoin Lite          } &   BTCL & 2018-01-31 \\
\textbf{Bitcoin Lunar         } &    BCL & 2018-03-20 \\
\textbf{Bitcoin Master        } &    BCM & 2018-03-24 \\
\textbf{Bitcoin Metal         } &   BTCM & 2018-05-01 \\
\textbf{Bitcoin Minor         } &    BTM & 2017-12-11 \\
\textbf{Bitcoin Nano          } &     BN & 2017-12-31 \\
\textbf{Bitcoin New           } &    BTN & 2017-12-25 \\
\textbf{Bitcoin Ore           } &    BCO & 2017-12-31 \\
\textbf{Bitcoin Parallel      } &    BCP & 2018-01-31 \\
\textbf{Bitcoin Pay           } &    BTP & 2017-12-15 \\
\textbf{Bitcoin Pizza         } &    BPA & 2017-12-31 \\
\textbf{Bitcoin Point         } &  POINT & 2017-12-25 \\
\textbf{Bitcoin Post-Quantum  } &    BPQ & 2018-12-22 \\
\textbf{Bitcoin Private       } &   BTCP & 2018-02-28 \\
\textbf{Bitcoin Pro           } &    BTP & 2018-01-31 \\
\textbf{Bitcoin Quantum       } &   QBTC & 2017-12-28 \\
\end{tabular}
\end{minipage} \hfill
\begin{minipage}{0.4\textwidth}
\fontsize{11pt}{11pt}\selectfont
\begin{tabular}{lll}
\toprule
\textbf{Fork name} & ticker &  Fork date \\
\midrule
\textbf{Bitcoin Reference     } &  BRECO & 2018-05-17 \\
\textbf{Bitcoin Rhodium       } &    XRC & 2018-01-10 \\
\textbf{Bitcoin RM            } &   BCRM & 2018-08-21 \\
\textbf{Bitcoin Smart         } &    BCS & 2018-01-19 \\
\textbf{Bitcoin Stake         } &   BTCS & 2017-12-18 \\
\textbf{Bitcoin Star          } &    BCS & 2018-01-07 \\
\textbf{Bitcoin Sudu          } &   SUDU & 2018-02-20 \\
\textbf{Bitcoin Top           } &    BTT & 2017-12-26 \\
\textbf{Bitcoin Transfer      } &   BTCT & 2018-04-01 \\
\textbf{Bitcoin Wonder        } &    BCW & 2017-12-18 \\
\textbf{Bitcoin World         } &    BTW & 2017-12-17 \\
\textbf{BitcoinX              } &    BCX & 2017-12-12 \\
\textbf{Bitcoinx2             } &  BTCX2 & 2018-07-01 \\
\textbf{Bitcoinzerox          } &    BZX & 2018-08-31 \\
\textbf{Bitcore               } &    BTX & 2017-11-02 \\
\textbf{Bitethereum           } &   BITE & 2017-12-21 \\
\textbf{Bithereum             } &    BTH & 2018-12-28 \\
\textbf{Bithereum             } &   BTH2 & 2018-12-28 \\
\textbf{Bitvote               } &    BTV & 2018-01-19 \\
\textbf{Cereneum              } &    CER & 2019-05-14 \\
\textbf{Clams                 } &   CLAM & 2014-05-12 \\
\textbf{Classicbitcoin        } &   CBTC & 2018-04-01 \\
\textbf{Dalilcoin             } &    DLC & 2015-03-30 \\
\textbf{Dash                  } &   DASH & 2014-01-18 \\
\textbf{Decred                } &    DCR & 2016-02-08 \\
\textbf{Digibyte              } &    DGB & 2014-01-10 \\
\textbf{Fastbitcoin           } &   FBTC & 2017-12-27 \\
\textbf{Fox BTC               } &   FBTC & 2018-04-30 \\
\textbf{Groestlcoin           } &    GRS & 2014-03-22 \\
\textbf{Lightning Bitcoin     } &   LBTC & 2017-12-18 \\
\textbf{Litecoin              } &    LTC & 2011-10-07 \\
\textbf{Microbitcoin          } &    MBC & 2018-05-28 \\
\textbf{Mimblewimblecoin      } &    MWC & 2019-07-19 \\
\textbf{Navcoin               } &    NAV & 2014-04-23 \\
\textbf{New Bitcoin           } &   NBTC & 2017-12-27 \\
\textbf{Oil Bitcoin           } &   OBTC & 2017-12-12 \\
\textbf{Qeditas               } &    QED & 2015-03-30 \\
\textbf{Smart Bitcoin         } &    SBC & 2018-04-20 \\
\textbf{Super Bitcoin         } &   SBTC & 2017-12-12 \\
\textbf{Syscoin               } &    SYS & 2014-07-19 \\
\textbf{Unitedbitcoin         } &   UBTC & 2017-12-12 \\
\textbf{Viacoin               } &    VIA & 2014-07-18 \\
\textbf{World Bitcoin         } &   WBTC & 2018-01-12 \\
\textbf{Xenon                 } &    XNN & 2018-06-30 \\
\textbf{Zcash                 } &    ZEC & 2016-10-28 \\
\bottomrule
\end{tabular}
\end{minipage}
\end{table}

\begin{table}[ht!]
     \caption{\label{tab:hrdf}\textbf{The list of Hard Forks} \\ \footnotesize{\textit{The list of all the hard forks considered in our study.}}}
    \begin{center}
   \resizebox{1\textwidth}{!}{
    \begin{tabular}{ccclccc}
\cline{1-3} \cline{5-7}
\textbf{Fork Name} &  \textbf{Symbol} &  \textbf{Fork Date} & \multicolumn{1}{c}{} & \textbf{Fork Name} & \textbf{Symbol} &  \textbf{Fork Date} \\ \cline{1-3} \cline{5-7} 
Bitcoin   Zero                            & BZX                                    & September 30,   2018                      &                      & Lightning Bitcoin                         & LBTC                                   & December 19,   2017                       \\
Micro Bitcoin                             & MBC                                    & May   30, 2018                            &                      & Bitcoin Faith                             & BTF                                    & December   19, 2017                       \\
ClassicBitcoin                            & CBTC                                   & April   1, 2018                           &                      & Bitcoin Pay                               & BTP                                    & December   15, 2017                       \\
Bitcoin Lite                              & BTCL                                   & November   17, 2017                       &                      & Super Bitcoin                             & SBTC                                   & August   12, 2017                         \\
Bitcoin Atom                              & BCA                                    & January   24, 2018                        &                      & Bitcoin Hot                               & BTH                                    & December   12, 2017                       \\
Bitcoin Interest                          & BCI                                    & January   22, 2018                        &                      & BitcoinX                                  & BCX                                    & December   12, 2017                       \\
BitVote                                   & BTV                                    & January   21, 2018                        &                      & UnitedBitcoin                             & UBTC                                   & December   12, 2017                       \\
Bitcoin Rhodium                           & XRC                                    & January   10, 2018                        &                      & Bitcoin Diamond                           & BCD                                    & November   24, 2017                       \\
Bitcoin Private                           & BTCP                                   & March   3, 2018                           &                      & Bitcore                                   & BTX                                    & November   2, 2017                        \\
Bitcoin God                               & GOD                                    & December   27, 2017                       &                      & Bitcoin Gold                              & BTG                                    & November   12, 2017                       \\
Bitcoin File                              & BIFI                                   & December   27, 2017                       &                      & Bitcoin Cash                              & BCH                                    & August   1, 2017                          \\ \cline{1-3} \cline{5-7} 
\end{tabular}
}
    \end{center}
\end{table}

\begin{table}[ht!]
     \caption{\label{tab:source}\textbf{Data extraction sources} \\ \footnotesize{\textit{Table summarizing the website visited to retrieve data and construct our dataset. The prices and volumes were recovered from CoinMarketCap, Royalton-crix and Spglobal. The information regarding the forks was retrieved from various websites, as shown below.}}}
    \begin{center}
    \resizebox{0.75\textwidth}{!}{
    \begin{tabular}{m{3.5cm}|m{9cm}}
      \textbf{Type of data} & \textbf{Source}\\
      \hline
      \text{Financial information} & \textit{https://coinmarketcap.com} \\
        & \textit{https://www.spglobal.com} \\
         & \textit{https://www.Royalton-crix.com} \\
      \hline
      \text{Fork related data} & \textit{www.forks.net} \\
       & \textit{https://coindar.org} \\
       & \textit{https://forkdrop.io} \\
       & \textit{https://cryptoli.st} \\
       & \textit{https://cryptoslate.com/} \\
       & \textit{https://miningpools.com/} \\
       & \textit{https://cryptocurrencyfacts.com/a-list-of-upcoming-bitcoin-forks-and-past-forks} \\
       & \textit{https://medium.com/@bithereumnetwork} \\
       & \textit{http://masterthecrypto.com} \\
       & \textit{https://masterthecrypto.com/breakdown-of-cryptocurrency-market} \\
       & \textit{https://unhashed.com/bitcoin-cryptocurrency-forks-list} \\
                & \textit{https://bitcointalk.org/}\\
          \hline
    \end{tabular}
    }
    \end{center}
\end{table}

\begin{table}[ht!]
\centering
\caption{\label{table:Model_Choice} \textbf{Information criterion and model choice} \\ \footnotesize{\textit{The table below shows the values of multiple Information Criteria for two models. The first model is the one described by Equation \ref{eq:egarch_var}. The second model is similar, but with the dummy variable $D(t)$ being replaced by another variable $C(t)$ that counts the number of forks occurring on each day. The findings indicate a slight preference for the initial model over the alternative model.}}}
\resizebox{0.4\textwidth}{!}{
\begin{tabular}{l|ll}
\textit{} & \textit{\textbf{$D(t)$}} & \textit{\textbf{$C(t)$}}\\
\hline
\textbf{Akaike}                 &   -5.0701                  & -5.0693  \\
\textbf{Bayes}             & -5.049                    & -5.0482\\
\textbf{Shibata}              & -5.0701                   & -5.0693 \\
\textbf{Hannan-Quinn}             & -5.0623                    & -5.0615                     \\              
\end{tabular}}
\end{table}

\end{document}